\documentclass[12pt]{article}
\usepackage[a4paper, margin=1in]{geometry}
\usepackage[T1]{fontenc}
\usepackage[utf8]{inputenc}
\usepackage{amsmath, amssymb, amsthm}
\usepackage{graphicx}
\usepackage{booktabs}
\usepackage{array}
\usepackage{placeins}
\usepackage{float}
\usepackage{tikz}
\usetikzlibrary{arrows.meta, positioning, fit}
\usepackage{setspace}
\usepackage{authblk}
\usepackage{natbib}
\usepackage{url}
\usepackage{doi}
\usepackage{hyperref}
\hypersetup{
    colorlinks=true,
    linkcolor=blue,
    citecolor=blue,
    urlcolor=blue
}

\newtheorem{proposition}{Proposition}

\title{Dynamic Amplification of Risk-Estimate Bias Through Differential Detection: A Markov Model for History-Based Covariates}

\author[1]{Hadar Sharvit\thanks{Corresponding author: \href{mailto:Hadar.Sharvit@mail.huji.ac.il}{Hadar.Sharvit@mail.huji.ac.il}}}
\author[1]{Micha Mandel}

\affil[1]{\small Department of Statistics and Data Science, The Hebrew University of
Jerusalem, Mount Scopus, Jerusalem 9190501, Israel}
\date{\today}

\begin{document}

\maketitle
\begin{center}
\small\textbf{Short title:} Differential Detection Bias in History-Based Covariates
\end{center}

\newpage

\begin{abstract}
History-based risk factors, such as known family history, recorded personal history, and documented regional history, are often treated as covariates. Because these variables depend on testing, diagnosis, and recording, their observed values are influenced by detection. However, monitoring data are often unavailable, making it difficult to distinguish true history effects from detection-driven associations.
We study the feedback loop that arises when observed history affects future monitoring and future monitoring affects which histories become observed. Prostate cancer screening serves as a case in point: knowing a family history may raise awareness and increase testing compared with having no known history. We develop Markov models for true and observed history states and show that differential detection can distort both observed risk ratios and the distribution of the observed history variable, and can create an apparent history effect even when true event risk does not depend on history. With history-dependent true risk, the observed history process is generally not Markovian, although the joint true-observed process is. Uniform incomplete detection can attenuate true history effects by shifting individuals with missed events into less recent observed-history states.
We also develop an inverse sensitivity-analysis framework that combines a published observed association, a baseline event probability, and plausible detection probabilities to obtain the implied true risk contrast. Numerical analyses illustrate the distortions, and a prostate cancer family-history example demonstrates the sensitivity calculation using external calibration values. The framework is intended for registry, electronic health record, and surveillance studies that use documented history as a risk factor.
\end{abstract}


\newpage

\section{Introduction}

Many epidemiological analyses use covariates that summarize past events. Examples include known family history of disease \citep{collaborative2001familial}, recorded personal history of previous episodes \citep{miller2007recent} and documented regional infection history, such as confirmed SARS-CoV-2 incidence in a city, region, or country \citep{ecdc2020testing}. These variables are not simply background characteristics. They are records of earlier events that had to be detected, reported, or remembered. The covariate available to the analyst is therefore often an error-prone observed history, not the underlying true history.

This distinction matters when observed history affects future monitoring. A detected event changes the recorded history, the recorded history may then change testing, surveillance, or clinical attention, and this altered monitoring affects which later events are detected. The result is a feedback mechanism in which the risk factor used in the analysis is partly generated by the detection process itself. In such settings, an observed association between history and disease may reflect true risk, differential detection, or both.

A typical example is prostate cancer screening. The American Cancer Society recommends that men at average risk discuss prostate cancer screening at age 50, while men at high risk, including those with a first-degree relative such as a father or brother diagnosed with prostate cancer before age 65, should have this discussion at age 45 \citep{acs2023prostate}. The AUA/SUO guideline recommends offering prostate cancer screening beginning at age 45 to 50 for people at average risk, and beginning at age 40 to 45 for people at increased risk, including those with strong family history \citep{wei2023aua}. Thus, known family history can change screening behavior. Such differences may arise not only from formal screening recommendations, but also from greater patient or clinician awareness following a known family diagnosis. At the same time, a relative's prostate cancer diagnosis may itself have depended on whether that relative was screened. The US Preventive Services Task Force notes that many men with prostate cancer would not experience symptoms and would not know they had the disease without screening \citep{uspstf2018}. Known family history is therefore not only an inherited or biological variable, it is also an observed diagnostic product.

The same structure appears in other individual-level history monitoring settings. For example, post-colonoscopy surveillance intervals are explicitly determined by previous colonoscopy findings, including the number, size, and histology of detected polyps \citep{gupta2020colonoscopy}. This example is more complicated than our basic model because colonoscopy may change future disease risk through polypectomy as well as future observation. COVID-19 provides two related examples. At the individual level, a previous recorded infection may affect the decision to test again. Early in the pandemic, reinfection was often considered unlikely in the short term \citep{mcmahon2020reinfection}. Under that belief, a documented previous infection could reduce the perceived need for repeat testing. This has the same feedback structure as our model, but the direction of the monitoring effect is reversed. At the regional level, a city, region, or country with a documented outbreak may become the target of more intensive testing. For example, ECDC guidance notes that population-wide testing may be appropriate in local areas with high incidence \citep{ecdc2020testing}. This creates an additional observation mechanism: the region has higher infection risk because there are more cases, but it may also have higher detection probability because more tests are performed.

A similar feedback mechanism appears outside medicine in predictive policing. \citet{ensign2018runaway} distinguish between reported incidents and discovered incidents, showing how allocating police according to previously discovered incidents can reinforce observed regional histories. The medical analogue is not identical, but the logic is similar: once a history is created by observation, using that history to guide future observation can reinforce detection bias.

These examples share a common feature: the history variable is not merely a record of past risk, but also a product of past observation. The issue is especially important for registry and electronic health records (EHR) analyses because the testing process is often incompletely observed. SEER incidence data \citep{seer2026data}, for example, are collected from population-based cancer registries and include information on diagnosed cancer cases such as demographics, tumor site, morphology, stage, treatment, and vital status. These data are essential for cancer surveillance, but they record diagnosed cancers rather than the underlying screening process. When no event is observed, we often cannot distinguish between no event, an event that was not tested, and a negative test that was not captured by the registry. Thus, a registry-only analysis usually cannot remove the problem simply by controlling for the testing indicator.

This paper focuses on observed-history covariates. Building on the diagnostic-feedback perspective of \citet{aikens2024feedback}, who study how biased diagnostic evidence can be reinforced in a one-step model, we extend the idea to repeated history-based monitoring. Past detection creates the recorded history, and recorded history guides future detection. We formulate this feedback as a Markov process, separating true history from observed history, and study its long-run consequences for risk ratios and for the population distribution of the history variable. We also show how the model can be used in reverse: published observed associations, external information on baseline risk, and plausible detection probabilities can be combined to obtain the true risk contrasts implied by different detection scenarios, with calibration uncertainty explored through sensitivity ranges. 

The model is intended primarily as a conceptual and sensitivity-analysis framework rather than as a generally identifiable correction procedure. In many applications, the detection process is only partially observed, so the true effect cannot be recovered without external information or assumptions about detection probabilities. Nevertheless, making this feedback mechanism explicit identifies a source of bias that may otherwise be overlooked and provides a structured way to assess how strongly plausible detection patterns could alter substantive conclusions.

The paper proceeds as follows. Section~\ref{sec:framework} introduces the observed-history framework and defines the true and observed history processes. Section~\ref{sec:fixed_detection} develops the main mathematical results. Section~\ref{sec:sensitivity_template} presents the inverse sensitivity-analysis template. Section~\ref{sec:simulation} uses simulations to study the size and direction of the bias across event probabilities and detection regimes. Section~\ref{sec:prostate_inverse} applies the method to a prostate cancer family-history calibration.

\section{Observed history as a detection-generated covariate}
\label{sec:framework}

We consider a discrete-time process over time intervals indexed by \(t\). An interval may represent a screening interval, a generation, or another unit appropriate to the application. Let $A_{t}$ denote an indicator of a true new event in interval $t$. For example, $A_{t}=1$ may represent the occurrence of cancer, a recurrent episode, or a regional infection event. Let $Z_{t}$ denote whether an event is detected and recorded, conditional on the event occurring. With no false positives, the indicator of an observed event is
\[
A^*_{t}=A_{t}Z_{t}.
\]
This representation absorbs several mechanisms into $Z_{t}$: clinical testing, access to care, coding, reporting, and communication of the event to those who use the history variable. False positives and imperfect test accuracy could be added by replacing this equation with a standard misclassification model, but in order to focus on the feedback mechanism, we ignore them.

A central theme is distinguishing true history from observed history. Let
\[
H_t\in\{1,2,\ldots,J+1\}
\]
denote the true history state at the end of interval \(t\). State \(1\) means that an event occurred during the interval ending at \(t\), that is, between \(t-1\) and \(t\). State \(2\) means that the most recent event occurred one interval earlier, and so on. The terminal state \(J+1\) collects all histories with no event during the last \(J\) intervals. Thus, \(H_t\) is determined after \(A_t\) has occurred. The update rule is

\[
H_t=
\begin{cases}
1, & A_t=1,\\
\min(H_{t-1}+1,J+1), & A_t=0.
\end{cases}
\]
Let \(H_t^*\) denote the observed history state, defined in the same way but using observed events:
\[
H_t^*=
\begin{cases}
1, & A_t^*=1,\\
\min(H_{t-1}^*+1,J+1), & A_t^*=0.
\end{cases}
\]

We assume no false positives and initialize the process with a compatible history, \(H_0\leq H_0^*\). Then \(H_t\leq H_t^*\) for all \(t\): the observed history cannot be more recent than the true history.

The event probability is allowed to depend on true history,
\begin{equation}
\label{eq:q_i}
q_i=\Pr(A_{t+1}=1\mid H_t=i),
\end{equation}
whereas the detection probability is allowed to depend on observed history,
\begin{equation}
\label{eq:p_j}
p_j=\Pr(Z_{t+1}=1\mid A_{t+1}=1,H_t^*=j).
\end{equation}
The parameters \(q_i\) and \(p_j\) are fixed state-specific parameters independent of $t$. An individual's event or detection probability may change as their history state changes, but the mapping from state to probability is fixed.

We only model \(Z_{t+1}\) conditional on \(A_{t+1}=1\). Under the no-false-positive assumption, $A_{t+1}^*=A_{t+1}Z_{t+1}$, 
so \(A_{t+1}^*=0\) whenever \(A_{t+1}=0\).

The distinction between \(q_i\) and \(p_j\) separates event occurrence from event detection. In the prostate example, \(H_t\) represents the true time since the last family prostate cancer event, while \(H_t^*\) represents the known or documented time since such an event. Known recent family history may increase prostate-specific antigen (PSA) screening through screening recommendations, clinical attention, or patient awareness, so \(p_1\) may exceed \(p_{J+1}\).

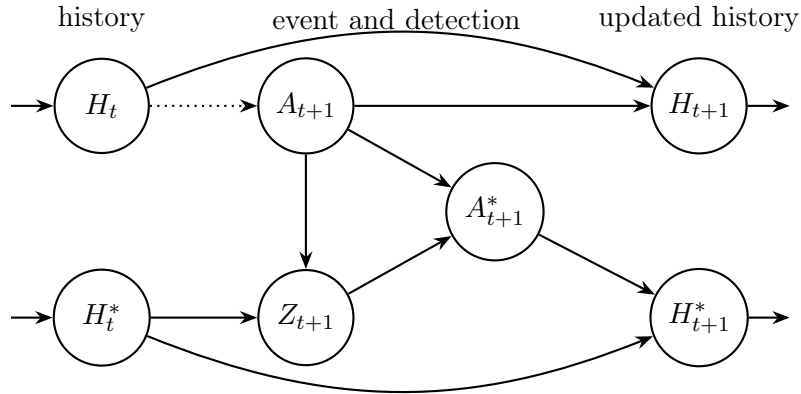
\begin{figure}[H]
\centering
\begin{tikzpicture}[>=Stealth, thick,
  circ/.style={draw, circle, minimum size=0.95cm, text width=0.82cm, align=center, font=\small}]

  \node[circ] (Ht) at (0,1.4) {$H_t$};
  \node[circ] (Hst) at (0,-1.4) {$H_t^*$};

  \node[circ] (A) at (2.7,1.4) {$A_{t+1}$};
  \node[circ] (Z) at (2.7,-1.4) {$Z_{t+1}$};
  \node[circ] (Ast) at (5.2,0) {$A^*_{t+1}$};

  \node[circ] (Hn) at (7.9,1.4) {$H_{t+1}$};
  \node[circ] (Hsn) at (7.9,-1.4) {$H^*_{t+1}$};

  \node[font=\small] at (0,2.55) {history};
  \node[font=\small] at (3.9,2.55) {event and detection};
  \node[font=\small] at (7.9,2.55) {updated history};

  \draw[dotted,->] (Ht) -- (A);
  \draw[->] (Hst) -- (Z);
  \draw[->] (A) -- (Ast);
   \draw[->] (A) -- (Z);
  \draw[->] (Z) -- (Ast);
  \draw[->] (A) -- (Hn);
  \draw[->] (Ht) to[bend left=22] (Hn);
  \draw[->] (Ast) -- (Hsn);
  \draw[->] (Hst) to[bend right=22] (Hsn);

  \draw[->] (-1.2,1.4) -- (Ht);
  \draw[->] (-1.2,-1.4) -- (Hst);

  \draw[->] (Hn) -- (9.1,1.4);
  \draw[->] (Hsn) -- (9.1,-1.4);

\end{tikzpicture}
\caption{History-based monitoring. True history affects the next event, while observed history affects detection. The detected event updates the observed history that guides future monitoring. The dotted arrow indicates that the effect of true history on the event process may be absent.}
\label{fig:history_dag}
\end{figure}

Figure~\ref{fig:history_dag} summarizes the feedback structure. The dotted arrow from $H_t$ to $A_{t+1}$ indicates that true history may or may not affect the underlying event process. In particular, one of our main concerns is the case in which the apparent effect of history arises entirely through differential detection, even when true event risk does not depend on history.

We study two important quantities. The first is the risk ratio comparing event probabilities across history states. This quantity is central because history-based variables are often used to define high-risk groups, guide monitoring decisions, and summarize associations in epidemiologic studies. The true risk ratio comparing history states \(i\) and \(j\) is
\[
RR_{ij}
=
\frac{\Pr(A_{t+1}=1\mid H_t=i)}
{\Pr(A_{t+1}=1\mid H_t=j)}
=
\frac{q_i}{q_j}.
\]
The corresponding observed risk ratio is
\begin{equation}
RR^*_{ij}
=
\frac{\Pr(A_{t+1}^*=1\mid H_t^*=i)}
{\Pr(A_{t+1}^*=1\mid H_t^*=j)}.
\label{eq:rr_observed_general}
\end{equation}
Thus, \(RR_{ij}\) is defined using the true history and true event, whereas \(RR^*_{ij}\) is defined using the observed history and observed event.

The second parameter we study is the stationary distribution of the observed history process \(H_t^*\), which we compare with the stationary distribution of the true history process \(H_t\). This comparison describes how the observation process changes the long-run composition of documented histories in the population. The stationary distribution is of interest because detection bias may change not only estimated associations, but also the apparent prevalence of the history-based risk factor itself.

\section{Risk ratios and history distributions under fixed detection}
\label{sec:fixed_detection}

Assume that \(q_i\) and \(p_j\) are fixed state-specific parameters. 

\subsection{Equal true risk}
\label{sec:equal_prevalence}

Suppose first that history is not a risk factor for the event, $q_1=q_2=\cdots=q_{J+1}=q$, and hence any observed association between history and diagnosis is due to the observation process.  The true risk ratio comparing any two history states is obviously
\[
RR_{ij}
=
\frac{\Pr(A_{t+1}=1\mid H_t=i)}
{\Pr(A_{t+1}=1\mid H_t=j)}
=
1.
\]
However, as detection depends on the observed history state, \(\Pr(A_{t+1}^*=1\mid H_t^*=j)=q p_j\), the observed risk ratio may differ from 1,
\begin{equation} 
RR^*_{ij}
=
\frac{\Pr(A_{t+1}^*=1\mid H_t^*=i)}
{\Pr(A_{t+1}^*=1\mid H_t^*=j)}
=
\frac{p_i}{p_j}.
\label{eq:rr_equal}
\end{equation}
Thus, if recent observed history increases monitoring, observed data may show an apparent history effect even when true risk is identical across history states.

This bias affects also the distribution of the putative risk factor in the population. The observed history process is a Markov chain that, for state \(j\), moves to state \(1\) with probability \(q p_j\) and move to state \(j+1\) (or remains in state $J+1$ for $j=J+1$) with probability $1-qp_j$.  

Let \(\pi_j^*\) be the stationary probability of observed history state \(j\). It satisfies the linear set of equations detailed in Appendix~\ref{app:equal}.

For adjacent non-terminal states, the observed stationary distribution satisfies
\[
\frac{\pi_j^*}{\pi_{j+1}^*}
=
\frac{1}{1-q p_j},
\qquad j=1,\ldots,J-1.
\]
Thus, if the detection probabilities are constant, \(p_j=p\), independent of the observed history, the adjacent ratios are also constant, and the observed history distribution is geometric over states \(1,\ldots,J\), with all remaining probability mass winsorized into the terminal state \(J+1\). Differential detection breaks this geometric structure: changes in \(p_j\) across history states change the adjacent ratios and therefore change the shape of the observed history distribution.

Therefore, in the equal-risk case, uniform detection is well behaved in two senses. It preserves the null observed risk ratio \eqref{eq:rr_equal}, \(RR^*_{ij}=1\), and it preserves the geometric form of the history distribution, although with parameter \(qp\) rather than \(q\). Detection focused on individuals with recent observed history has the opposite effect: it creates a non-null observed risk ratio and also produces a non-geometric observed history distribution. Both features may therefore look like evidence of a history effect even when the true event probability is the same in all history states. 

Shape is only one aspect of the distributional distortion. We also compare the true and observed stationary distributions directly using

\begin{equation}
D_H
=
\sum_{j=1}^{J+1}\{F_H(j)-F_{H^*}(j)\}
=
E(H^*)-E(H),
\label{eq:distortion}
\end{equation}
where  \(F_H\) and \(F_{H^*}\) are the stationary cumulative distribution functions of the true and observed histories. The representation as a difference of expectations between $H^*\sim F_{H^*}$ and $H\sim F_{H}$ holds because \(H^*\) and \(H\) are integer-valued variables with the same ordered support \(\{1,\ldots,J+1\}\). Since \(H\leq H^*\), also \(D_H\geq0\), and  the measure is useful when comparing testing regimes. 

The size of this discrepancy depends on \(q\), the full detection profile \(p_1,\ldots,p_{J+1}\), and \(J\). When \(q\) is small, a large stationary mass lies in the terminal state \(J+1\), so a detection rule that concentrates testing among recent observed histories may miss many events among those with no known history and push additional mass into the terminal state. When \(q\) is larger, more mass is naturally located in recent states, so the same detection profile may have a different effect. For this reason, the ordering of detection regimes by distributional accuracy depends on the event probability and the full detection profile. We demonstrate this in the simulation study below.

\subsection{History-dependent true risk}
\label{sec:history_dependent}

We next consider the model in which history is a true risk factor, so that the event probability \eqref{eq:q_i} is a function of true history, while the detection probability \eqref{eq:p_j} remains a function of observed history.

The true risk ratio between history states \(i\) and \(j\) is $RR_{ij}=q_i/q_j$. To find the observed risk ratio, we first note that the observed probability of the event among people with observed history \(j\) is
\begin{equation}
\Pr(A_{t+1}^*=1\mid H_t^*=j)
= p_j\sum_{k=1}^{j}q_k\Pr(H_t=k\mid H_t^*=j).
\label{eq:observed_event_general}
\end{equation}
Therefore,
\begin{equation}
RR^*_{ij}=\frac{p_i\sum_{k=1}^{i}q_k\Pr(H_t=k\mid H_t^*=i)}
{p_j\sum_{k=1}^{j}q_k\Pr(H_t=k\mid H_t^*=j)}.
\label{eq:observed_rr_general}
\end{equation}
This expression separates two sources of bias. The ratio \(p_i/p_j\) reflects differential detection, and the conditional mixtures \(\Pr(H_t=k\mid H_t^*=j)\) reflect hidden true histories inside observed-history groups.

Here, the process $H_t^*$ is generally not Markovian, as people having the same observed history can have different true histories, and these true histories have different probabilities of experiencing the event. Nevertheless, the pair $(H_t,H_t^*)$ is Markovian on
\[
\mathcal S=\{(i,j):1\leq i\leq j\leq J+1\},
\]
with transition probabilities 
\begin{equation}
P_{(i,j)\to(k,\ell)}=
\begin{cases}
q_i p_j, & (k,\ell)=(1,1),\\[4pt]
q_i(1-p_j), & (k,\ell)=\left(1,\min(j+1,J+1)\right),\\[4pt]
1-q_i, & (k,\ell)=\left(\min(i+1,J+1),\min(j+1,J+1)\right),\\[4pt]
0, & \text{otherwise.}
\end{cases}
\label{eq:joint_transition}
\end{equation}
Appendix~\ref{app:unequal} describes the set of linear equations that defines the stationary distribution of the joint chain. Interestingly, it is shown in Appendix~\ref{app:unequal} that the stationary probability of the terminal state \((J+1,J+1)\) is determined entirely by the true event process and does not depend on the detection probabilities. 

\subsection{The two-state case}
\label{sec:j1}

It is easier to explore the bias in the simplest and most important two-state case, \(J=1\). In this case, the history variable records whether an event occurred in the previous interval. This case is also the closest analogue to many applications in which history is recorded as a binary indicator, such as reported first-degree family history versus no reported first-degree family history. The joint process has three possible states:
state $(1,1)$ of recent true history and recent observed history, state $(1,2)$ of recent true history that was not observed, and state $(2,2)$ of neither true nor observed recent history. With rows and columns ordered lexicographically, the transition matrix is
\[
M=
\begin{pmatrix}
q_1p_1 & q_1(1-p_1) & 1-q_1\\
q_1p_2 & q_1(1-p_2) & 1-q_1\\
q_2p_2 & q_2(1-p_2) & 1-q_2
\end{pmatrix}.
\]
Solving $\pi M=\pi$ gives
\[
\begin{aligned}
\pi_{22} &= \frac{1-q_1}{1-q_1+q_2},\\[4pt]
\pi_{12} &= \frac{q_2\{1-q_1p_1-p_2(1-q_1)\}}
{(1-q_1+q_2)\{1+q_1(p_2-p_1)\}},\\[4pt]
\pi_{11} &= \frac{q_2p_2}
{(1-q_1+q_2)\{1+q_1(p_2-p_1)\}}.
\end{aligned}
\]
The derivation is given in Appendix~\ref{app:J1}. As discussed above for the general case, the stationary probability of state $(2,2)$ depends only on the parameters $q_1$ and $q_2$ of the true history process. Detection probabilities split the recent true-history state into observed and hidden components.

The true risk ratio comparing individuals with and without recent  history  is
\[
RR=\frac{q_1}{q_2}.
\]
The stationary observed risk ratio compares the probabilities of a detected event among those with and without observed recent history:

\begin{equation}
RR^*=
RR\times \frac{q_2p_1}
{p_2\{q_1\pi_{12}+q_2\pi_{22}\}/(\pi_{12}+\pi_{22})}.
\label{eq:j1_rr}
\end{equation}

In the two-state model, the distributional discrepancy is also especially simple. The only discordant joint state is \((1,2)\): recent true history that is not observed as recent. Therefore
\[
D_H=E(H_t^*)-E(H_t)=\pi_{12}.
\]

\medskip
\noindent

Assuming
\begin{equation}
\label{eq:qp_conditions}
0<q_2\leq q_1<1,
\qquad
0<p_2\leq p_1\leq 1,
\end{equation}
that is, true history may increase event risk and observed history may increase detection, the model satisfies the following proposition.

\begin{proposition}
\label{prop:j1_rr}
In the \(J=1\) case under \eqref{eq:qp_conditions}:
\begin{enumerate}
\item If \(p_1=p_2=p<1\), then \(RR^*\leq RR\), with equality only if \(p=1\) or \(q_1=q_2\).
\item If \(p_1=1\) and \(p_2<1\), then \(RR^*>RR\).
\item For fixed \(q_1,q_2\), and \(p_2\), \(RR^*\) is increasing in \(p_1\).
\item If \(p_2<1\) and \(q_1>q_2\), there is a unique value \(p_1^*\in(p_2,1)\) such that \(RR^*=RR\). This threshold is given explicitly in Appendix~\ref{app:RRJ1}.
\item \(\pi_{11}\) is increasing in both \(p_1\) and \(p_2\).
\end{enumerate}
\end{proposition}

The proof is given in Appendix~\ref{app:RRJ1}.

The proposition shows how incomplete and differential detection affect the observed risk ratio. Parts 1 and 2 give the two extremes. Under uniform incomplete detection, \(RR^*\) is attenuated relative to \(RR\), because the reference group contains individuals with recent true history that was not observed. At the other extreme, if detection is complete among those with observed recent history but incomplete in the reference group, then \(RR^*\) exceeds \(RR\). Part 3 shows that, between these extremes, increasing \(p_1\) moves \(RR^*\) upward. Together with continuity, this gives the unique threshold in Part 4: there is one value of \(p_1\) at which the attenuation from hidden true history exactly balances the upward bias from differential detection.

Part 5 connects this risk-ratio behavior to the observed-history distribution. In the two-state model, recent true history is split between observed recent history, \((1,1)\), and hidden recent history, \((1,2)\). Thus, for fixed \(q_1\) and \(q_2\), increasing \(\pi_{11}\) reduces \(\pi_{12}=D_H\). But improving agreement between true and observed histories is not the same as improving risk-ratio interpretation: increasing \(p_1\) can reduce hidden history while also moving \(RR^*\) above \(RR\). Thus, a monitoring rule can improve the observed-history distribution while worsening relative-risk interpretation.

\section{Sensitivity-analysis}
\label{sec:sensitivity_template}

The functional relationship between the model parameters and the stationary distribution provides a template for sensitivity analysis in settings where individual-level testing data are unavailable. An analyst may have an estimate of the observed association between the putative risk factor and the outcome from the literature, external information on the baseline event probability, and plausible assumptions about detection probabilities. The Markov model then determines the true risk contrast implied by each detection scenario.

Consider first the two-state case, \(J=1\). Suppose an observed risk ratio $\widehat{{RR}^{*}}$ is available from a registry, electronic health record study, or published analysis. Suppose also that \(q_2\), the baseline true event probability among individuals with no recent true history, and \(p_2\), the detection probability among individuals with no recent observed history, can be calibrated from external sources. For each sensitivity value of \(p_1\), the detection probability among individuals with recent observed history, we solve
\[
RR^*(q_1,q_2,p_1,p_2)
=
\widehat{RR^*}
\]
for \(q_1\), where \(RR^*(q_1,q_2,p_1,p_2)\) is given by Equation~\eqref{eq:j1_rr}. 
The solution, \(q_1(p_1)\), gives the true event probability \(q_1\) implied by the observed association under that detection scenario. The corresponding implied true risk ratio is
\[
RR_{\mathrm{impl}}(p_1)
=
\frac{q_1(p_1)}{q_2}.
\]
If no admissible solution exists, with \(0<q_1<1\), then the chosen detection scenario is incompatible with the observed risk ratio and the baseline calibration.

This inverse calculation provides a practical framework for sensitivity analysis. Rather than asking only how detection bias can distort an observed association, it asks how large the true history effect must be in order to produce the observed association under specified detection assumptions. Varying \(p_1\) over a plausible range produces a curve of implied true risk ratios. If \(p_1=p_2\), the calculation corresponds to uniform incomplete detection. If \(p_1>p_2\), the calculation allows for targeted detection among individuals with recent observed history.

The same idea extends to more than two history states. For example, in a three-state model with reference state \(3\), suppose the analyst has two observed risk ratios,
\[
\widehat{RR}^{*}_{13}
\qquad\text{and}\qquad
\widehat{RR}^{*}_{23}.
\]
If \(q_3\) and \(p_3\) are externally calibrated, then for each pair of sensitivity values \((p_1,p_2)\), the analyst solves the system
\[
RR^*_{13}(q_1,q_2,q_3,p_1,p_2,p_3)
=
\widehat{RR}^{*}_{13},
\]
\[
RR^*_{23}(q_1,q_2,q_3,p_1,p_2,p_3)
=
\widehat{RR}^{*}_{23},
\]
where the observed risk ratios are computed from the stationary distribution of the joint chain using Equation~\eqref{eq:observed_rr_general}. The unknowns are \(q_1\) and \(q_2\). The output is therefore a pair of implied true ratios,
\[
\frac{q_1}{q_3}
\qquad\text{and}\qquad
\frac{q_2}{q_3},
\]
for each assumed detection regime. These values can be displayed as curves when one detection parameter is varied at a time, or as heatmaps over a grid of \((p_1,p_2)\).

Uncertainty in the observed association can be incorporated, but the transformation should be treated with care. If the reported confidence interval for the observed association is
\[
[L^*,U^*],
\]
then, for each fixed detection scenario, the implied values of \(q_1\) can be defined as the set of admissible solutions satisfying
\[
L^*
\leq
RR^*(q_1,q_2,p_1,p_2)
\leq
U^*,
\qquad 0<q_1<1.
\]

The corresponding conditional interval on the true-risk-ratio scale is obtained by taking the range of \(q_1/q_2\) over this admissible solution set. This interval is conditional on the chosen values of \(q_2\), \(p_2\), and \(p_1\), and therefore reflects only the uncertainty in the observed association under those fixed calibration and detection assumptions.

In practice, the calibration inputs \(q_2\) and \(p_2\) may also be uncertain. The same inverse calculation can then be repeated over plausible values of these inputs. When several uncertain inputs are involved, this can be implemented as a Monte Carlo sensitivity analysis. When only one or two calibration inputs are varied, a deterministic grid may be clearer and more reproducible. In either case, the observed association and its reported limits can be held fixed, and the inverse calculation can be applied to the point estimate and to the full admissible set defined by the reported interval.

Thus, the deterministic inverse calculation provides the implied true ratio under fixed assumptions. In section \ref{sec:prostate_inverse} we demonstrate the sensitivity calculation using external calibration values for prostate cancer family history.

\section{Simulation study of detection regimes}
\label{sec:simulation}

This section presents a simulation study illustrating the risk-bias feedback-loop mechanisms implied by the Markov model. It examines the magnitude of bias that differential detection can create, attenuate, or amplify in \(RR^*\), and how different detection regimes can shift the observed history distribution away from the true distribution.

We first set \(J=4\), giving five history states. State \(1\) denotes the most recent history and state \(5\) denotes no history within the window. This is richer than many real family-history variables, which are often closer to binary indicators such as known first-degree family history. We use five states to make the distributional consequences of missed events visible across several history ages. The two-state case is used at the end of this section in order to study more directly the tradeoff between risk-ratio accuracy and distributional accuracy.

Throughout this section, we focus on the risk ratio between the most recent true-history state and the reference group of no history in the past four intervals,
\[
RR_{1,J+1}=\frac{q_1}{q_{J+1}}.
\]
The observed analogue is
\[
RR^*_{1,J+1}
=
\frac{\Pr(A_{t+1}^*=1\mid H_t^*=1)}
{\Pr(A_{t+1}^*=1\mid H_t^*=J+1)}.
\]
For readability, we write \(RR,RR^*\) for \(RR_{1,J+1},RR^*_{1,J+1}\) in the tables and figures.

For each scenario, we construct the joint transition matrix of \((H,H^*)\), compute its stationary distribution, and then calculate the reported summaries: the true and observed risk-ratios $RR$ and $RR^*$,
the distributional discrepancy \(D_H=E(H^*)-E(H),\)
and the recent-history coverage,
\[
C_1=\frac{\Pr(H^*=1)}{\Pr(H=1)}.
\]
With no-false-positives, \(H^*=1\) implies \(H=1\), so $C_1=\Pr(H^*=1\mid H=1).$
Thus, \(C_1\) is the stationary probability that a true recent history is represented as recent observed history.

We consider three equal-risk settings:
\[
\begin{array}{ll}
\text{Rare event, equal risk:} & q=(0.01,0.01,0.01,0.01,0.01),\\
\text{Common event, equal risk:} & q=(0.29,0.29,0.29,0.29,0.29),\\
\text{High-frequency event, equal risk:} & q=(0.50,0.50,0.50,0.50,0.50).
\end{array}
\]
The rare setting represents events that are uncommon in the underlying process. The common and high-frequency settings are stylized high-event-frequency regimes. The value \(q=0.29\) is motivated by the latent-prevalence scale in autopsy evidence on prostate cancer \citep{jahn2015autopsy}. Similarly, the detection profiles below are generic sensitivity regimes rather than estimates of prostate cancer screening probabilities.

We also consider three settings in which risk decreases with past history:
\[
\begin{array}{ll}
\text{Rare event, history effect:} & q=(0.030,0.024,0.019,0.015,0.010),\\
\text{Common event, history effect:} & q=(0.450,0.390,0.350,0.320,0.290),\\
\text{High-frequency event, history effect:} & q=(0.600,0.570,0.540,0.520,0.500).
\end{array}
\]
The history-effect settings encode higher true risk after more recent true history. The rare-event setting has a larger relative gradient, while the common and high-frequency settings keep the probabilities in a high latent-event range.

For the detection probabilities, complete observation is the benchmark; when \(p_j=1\) for all \(j\), \(RR^*=RR\), \(D_H=0\), and \(C_1=1\). We therefore focus on four incomplete detection regimes:
\[
\begin{array}{ll}
\text{Uniform incomplete:} & p=(0.35,0.35,0.35,0.35,0.35),\\
\text{Weak heterogeneity:} & p=(0.40,0.375,0.35,0.325,0.30),\\
\text{Moderate heterogeneity:} & p=(0.65,0.45,0.30,0.22,0.13),\\
\text{Strong heterogeneity:} & p=(0.85,0.50,0.25,0.10,0.05).
\end{array}
\]
The four regimes have the same arithmetic mean detection probability, \(0.35\). Thus, they differ in the testing-probability gradients across the five groups rather than in the simple average of the specified probabilities. These values are chosen to compare different gradients while holding the unweighted mean of the state-specific detection probabilities fixed.

Table~\ref{tab:simulation_equal} reports the equal-risk scenarios, in which all risk-ratio bias is due to detection. Uniform incomplete detection gives \(RR^*=1\), because detection does not depend on observed history. Nevertheless, the observed history distribution is distorted: \(D_H>0\) and \(C_1<1\), because detection is incomplete. Heterogeneity in detection probabilities creates an apparent association with recent history even though \(RR=1\). The stronger the heterogeneity, the larger the observed risk-ratio distortion and the larger the shift in the observed history distribution.

\begin{table}[tbp]
\centering
\caption{Simulation results under equal true risk.}
\label{tab:simulation_equal}
\small
\resizebox{\textwidth}{!}{%

\begin{tabular}{llrrrr}
\toprule
True risk setting & Detection regime & $RR$ & $RR^*$ & $D_H$ & $R_1$\\
\midrule
Rare event, q=0.01 & Uniform & 1 & 1.000 & 0.064 & 0.350\\
Rare event, q=0.01 & Weak & 1 & 1.333 & 0.069 & 0.301\\
Rare event, q=0.01 & Moderate & 1 & 5.000 & 0.086 & 0.131\\
Rare event, q=0.01 & Strong & 1 & 17.000 & 0.094 & 0.051\\
\addlinespace
Common event, q=0.29 & Uniform & 1 & 1.000 & 1.257 & 0.350\\
Common event, q=0.29 & Weak & 1 & 1.333 & 1.343 & 0.321\\
Common event, q=0.29 & Moderate & 1 & 5.000 & 1.731 & 0.180\\
Common event, q=0.29 & Strong & 1 & 17.000 & 1.984 & 0.080\\
\addlinespace
High-frequency event, q=0.50 & Uniform & 1 & 1.000 & 1.593 & 0.350\\
High-frequency event, q=0.50 & Weak & 1 & 1.333 & 1.685 & 0.334\\
High-frequency event, q=0.50 & Moderate & 1 & 5.000 & 2.202 & 0.229\\
High-frequency event, q=0.50 & Strong & 1 & 17.000 & 2.646 & 0.120\\
\bottomrule
\end{tabular}

}
\end{table}

When history is a true risk factor (\(RR>1\)), the results are more mixed, as shown in Table~\ref{tab:simulation_history}. Uniform incomplete detection attenuates \(RR^*\), because the reference group \(J+1\) contains individuals with hidden histories of the disease. Weak heterogeneity can sometimes offset this attenuation and improve risk-ratio accuracy, but it does not necessarily improve the observed history distribution. For example, in the common-event history-effect setting, weak heterogeneity gives \(RR^*=1.650\), which is closer to the true value \(RR=1.552\) than the value under uniform incomplete detection, but it also gives larger \(D_H\) and smaller \(C_1\). Stronger heterogeneity can be much more problematic: even in the rare-event setting, where the true risk ratio is \(RR=3\), moderate and strong heterogeneity inflate the observed risk ratio to \(14.388\) and \(48.732\), respectively, while also increasing \(D_H\) and reducing \(C_1\). Even when history is a true risk factor, the observed risk ratio can be biased upward or downward depending on the detection regime.

\begin{table}[tbp]
\centering
\caption{Simulation results under history-dependent true risk.}
\label{tab:simulation_history}
\small
\resizebox{\textwidth}{!}{%

\begin{tabular}{llrrrr}
\toprule
True risk setting & Detection regime & $RR$ & $RR^*$ & $D_H$ & $R_1$\\
\midrule
Rare event & Uniform & 3.000 & 2.908 & 0.066 & 0.350\\
Rare event & Weak & 3.000 & 3.868 & 0.071 & 0.302\\
Rare event & Moderate & 3.000 & 14.388 & 0.088 & 0.134\\
Rare event & Strong & 3.000 & 48.732 & 0.097 & 0.052\\
\addlinespace
Common event & Uniform & 1.552 & 1.248 & 1.351 & 0.350\\
Common event & Weak & 1.552 & 1.650 & 1.436 & 0.329\\
Common event & Moderate & 1.552 & 6.011 & 1.875 & 0.210\\
Common event & Strong & 1.552 & 20.166 & 2.213 & 0.103\\
\addlinespace
High-frequency event & Uniform & 1.200 & 1.055 & 1.609 & 0.350\\
High-frequency event & Weak & 1.200 & 1.402 & 1.693 & 0.340\\
High-frequency event & Moderate & 1.200 & 5.213 & 2.211 & 0.254\\
High-frequency event & Strong & 1.200 & 17.655 & 2.713 & 0.145\\
\bottomrule
\end{tabular}

}
\end{table}

\begin{figure}[tbp]
\centering
\includegraphics[width=0.95\textwidth]{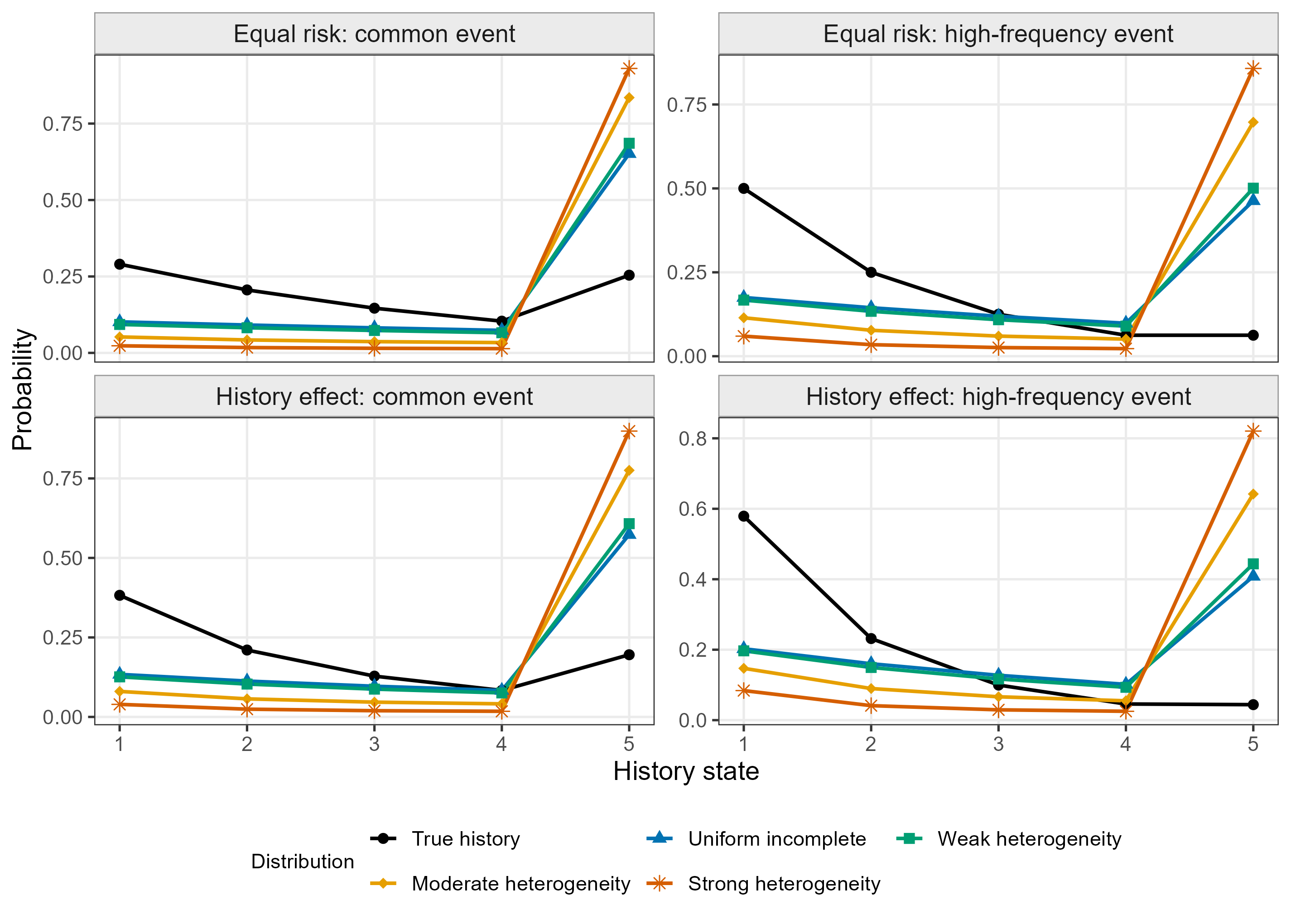}
\caption{Stationary true and observed history distributions under equal true risk and history-dependent true risk.}
\label{fig:history_distributions_combined}
\end{figure}

Figure~\ref{fig:history_distributions_combined} visualizes the corresponding stationary history distributions for the common-event and high-frequency-event settings. It complements Tables~\ref{tab:simulation_equal} and~\ref{tab:simulation_history} by showing where the distributional discrepancies summarized by \(D_H\) and \(C_1\) occur across the history states.

The simulations also show that concentrating detection among recent observed-history states is not automatically beneficial. In Tables~\ref{tab:simulation_equal} and~\ref{tab:simulation_history}, the strong heterogeneity regime performs worse than uniform incomplete detection in all event-frequency settings on the main summaries: it produces larger risk-ratio distortion, larger \(D_H\), and smaller \(C_1\). This is especially visible in the common-event and high-frequency-event settings, where the stationary distributions in Figure~\ref{fig:history_distributions_combined} show larger shifts of observed history away from true history under strong heterogeneity. The reason is that this regime concentrates testing among individuals with recent observed history. When many true events occur among individuals with no recent observed history, this strategy misses cases from the larger reference pool and pushes additional mass into less recent observed-history states.

\begin{figure}[H]
\centering
\includegraphics[width=0.86\textwidth]{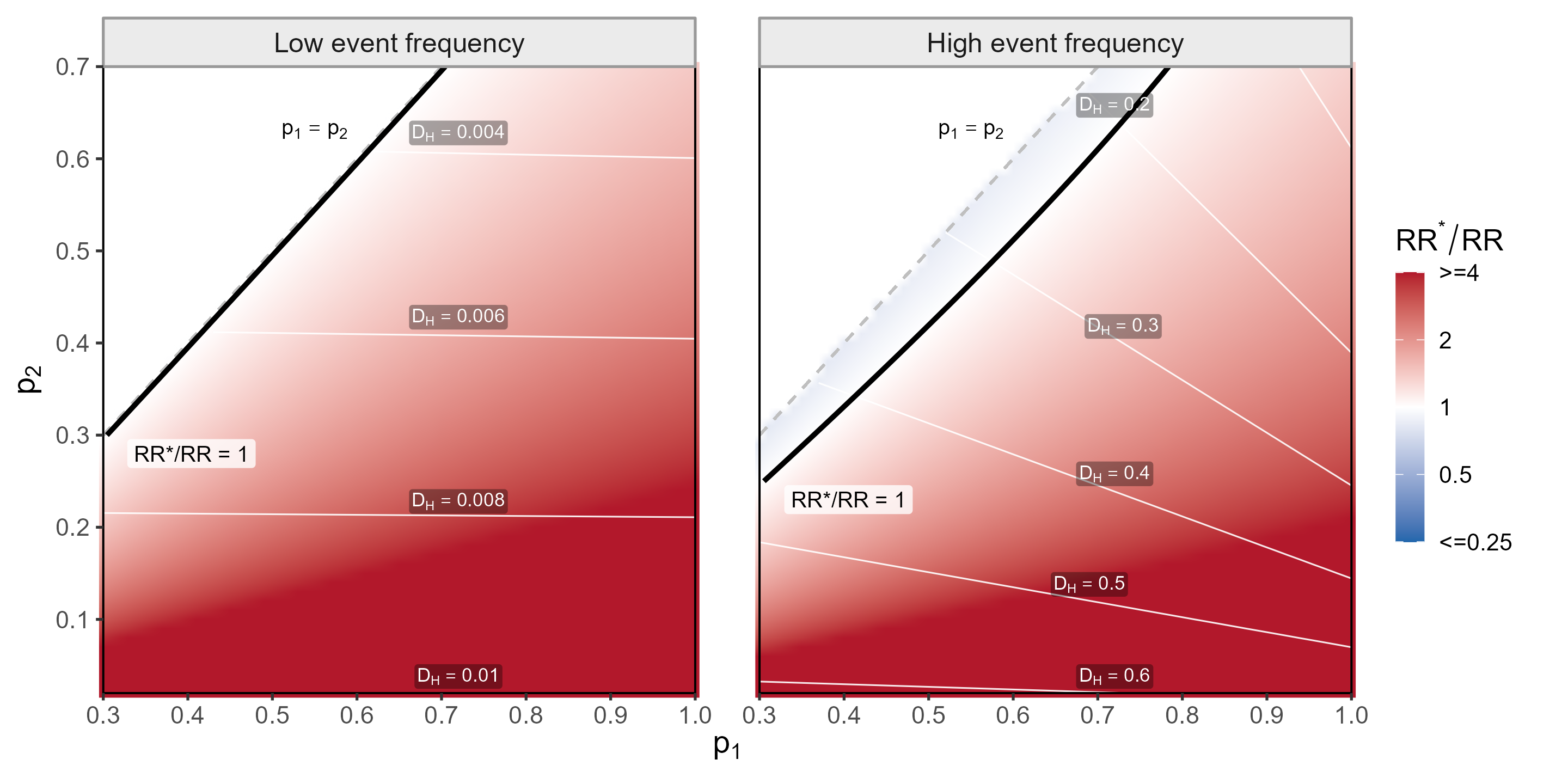}
\caption{Two-state tradeoff between risk-ratio accuracy and distributional accuracy. The left panel shows a low-event-frequency setting, \(q_1=0.03\) and \(q_2=0.01\), and the right panel shows a high-event-frequency setting, \(q_1=0.70\) and \(q_2=0.50\). Color shows \(RR^*/RR\): values below one indicate underestimation and values above one indicate overestimation of the true risk ratio. White contours show \(D_H\), the black curve marks \(RR^*/RR=1\), and the dashed diagonal marks uniform detection, \(p_1=p_2\). In the low-event-frequency setting, the \(RR^*/RR=1\) curve lies close to the uniform-detection line, while in the high-event-frequency setting the conflict between reducing \(D_H\) and preserving \(RR^*/RR\) is more pronounced.}
\label{fig:j1_tradeoff}
\end{figure}

Finally, Figure~\ref{fig:j1_tradeoff} returns to the two-state model and compares two event-frequency settings. The left panel uses a low-event-frequency setting, \(q_1=0.03\) and \(q_2=0.01\). The right panel uses a stylized high-event-frequency setting, \(q_1=0.70\) and \(q_2=0.50\). These high values are not meant to represent prostate cancer probabilities. They are used to show a setting in which the tension between distributional accuracy and risk-ratio accuracy becomes more visible.

In both panels, the horizontal axis is \(p_1\), the detection probability for individuals with observed recent history, and the vertical axis is \(p_2\), the detection probability for individuals with no observed recent history. Each point represents one detection regime \((p_1,p_2)\). The color shows \(RR^*/RR\), with values below one indicating underestimation of the true risk ratio and values above one indicating overestimation. The black curve marks \(RR^*/RR=1\), the white contours show \(D_H\), and the dashed diagonal marks uniform detection, \(p_1=p_2\).

The two panels show different relations between the two targets. In the low-event-frequency setting, the black curve is close to the diagonal \(p_1=p_2\), so risk-ratio accuracy is achieved approximately under uniform detection. In this panel, increasing \(p_1\) has little effect on \(D_H\), but it can still move \(RR^*\) away from \(RR\). Thus, differential detection can distort the observed risk ratio even when it does little to improve the observed-history distribution. In the high-event-frequency setting, the tradeoff is stronger. Increasing detection can reduce \(D_H\), but moving toward a more accurate observed-history distribution may also move the regime away from the black curve and distort \(RR^*/RR\). Hence distributional accuracy and risk-ratio accuracy need not be optimized by the same detection regime.

\section{Sensitivity analysis: a prostate cancer illustration}
\label{sec:prostate_inverse}

We apply the sensitivity-analysis template to prostate cancer family history. The aim is not to estimate prostate cancer natural history, but to show how published associations and external calibration values can be translated into implied true risk contrasts under different detection assumptions.

We begin with the baseline calibration. For the event probability among men with no known family history, we use
\[
q_2\in\{0.02,0.03,0.04\}.
\]
The central value \(q_2=0.03\) is based on the autopsy estimate of \(29\%\) latent prostate cancer among men aged 60-69 reported by \citet{jahn2015autopsy}, used here only to motivate a stylized annual probability for a first prostate cancer event over this age range. For the baseline detection probability, we use \(p_2=0.38\), based on the reported annual PSA-testing rate among men aged 55-69 as a proxy for detection \citep{nci2025ctpr}.

The published association is taken from \citet{lesko1996family}, who reported increased prostate cancer risk among men who reported a history of prostate cancer in either their fathers or brothers. We treat this reported family history as the observed-history variable. The original study reports an odds ratio of \(2.3\), with \(95\%\) confidence interval \([1.7,3.3]\). Because the outcome probability in the reference group is small under our calibration, the odds ratio is only slightly larger than the corresponding risk ratio. We therefore use
\[
RR^*_{\mathrm{obs}}=2.3,
\qquad
[L^*,U^*]=[1.7,3.3]
\]
as the observed association for the inverse calculation.

These inputs are calibration values rather than jointly estimated parameters. They come from different studies, populations, and time periods, and we use them together only to illustrate the sensitivity-analysis calculation. The resulting values should therefore be interpreted as literature-based sensitivity results, not as estimates from a single coherent prostate cancer cohort.

The sensitivity parameter is \(p_1\), the detection probability among men with observed family history. To choose a reference value, we use external evidence on differential PSA testing. \citet{drake2008screening} reported that men with a family history of prostate cancer were more likely to have ever had a PSA test, with an odds ratio of about \(1.8\) compared with men without family history. This estimate is not a direct estimate of \(p_1/p_2\), but it provides a plausible calibration for differential screening. Starting from \(p_2=0.38\), converting the odds ratio to the probability scale gives \(p_1\approx0.52\).

For each value of \(p_1\), and for each value of
\(q_2\in\{0.02,0.03,0.04\}\), we evaluate
Equation~\eqref{eq:j1_rr} over a fine grid of \(q_1\) values satisfying
\[
q_2\leq q_1\leq \min(10q_2,0.30).
\]
The point estimate is the minimal-effect grid value corresponding most closely
to \(RR^*_{\mathrm{obs}}=2.3\). The conditional interval is obtained by
taking the range of \(q_1/q_2\) over all grid values for which
\(1.7\leq RR^*\leq3.3\).

\begin{figure}[H]
\centering
\includegraphics[width=0.78\textwidth]{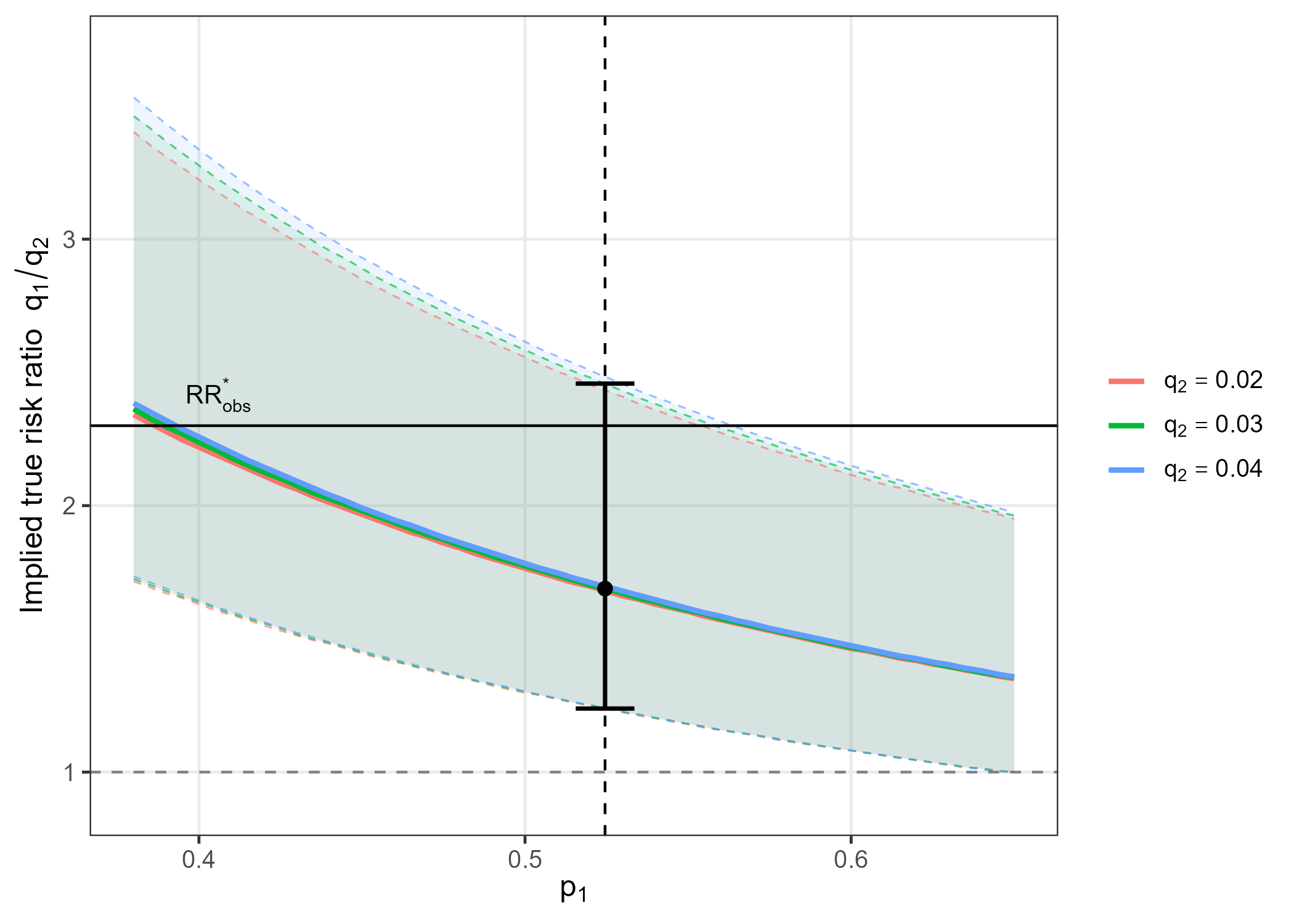}
\caption{Inverse sensitivity analysis under the two-state model. The curves show the implied true risk ratio \(q_1/q_2\) for different baseline values of \(q_2\). Shaded bands show the conditional intervals obtained from the admissible grid values for which \(1.7\le RR^*\le3.3\). The horizontal line marks the observed association, \(RR^*_{\mathrm{obs}}=2.3\). The vertical dashed line marks \(p_1\approx0.52\), corresponding to the empirical PSA-testing gradient. The black point and interval show the implied estimate and conditional interval for \(q_2=0.03\) at this empirical value of \(p_1\).}
\label{fig:prostate_inverse}
\end{figure}

At the left edge of the figure, where \(p_1=p_2=0.38\), detection is uniform and incomplete, and the observed risk ratio is smaller than the true risk ratio. As \(p_1\) increases, differential detection explains more of the observed association, and the implied true risk ratio declines. At the empirical screening-gradient value \(p_1\approx0.52\), with \(q_2=0.03\) and \(p_2=0.38\), the implied true risk ratio is approximately \(1.69\), with conditional interval \([1.24,2.46]\). Thus, under this detection scenario, the observed association is compatible with a substantially smaller true family-history effect.

As a secondary analysis, we repeat the inverse calculation while varying \(q_2\) and \(p_2\), which in the previous analysis were taken from external sources rather than from the same population as the observed association, thus may underestimate uncertainty. As there are only two inputs (\(q_2\) and \(p_2\)), with the observed $RR$  kept fixed at its point estimate, we use a deterministic grid over the calibration region rather than random sampling.

\begin{figure}[tbp]
\centering
\includegraphics[width=0.78\textwidth]{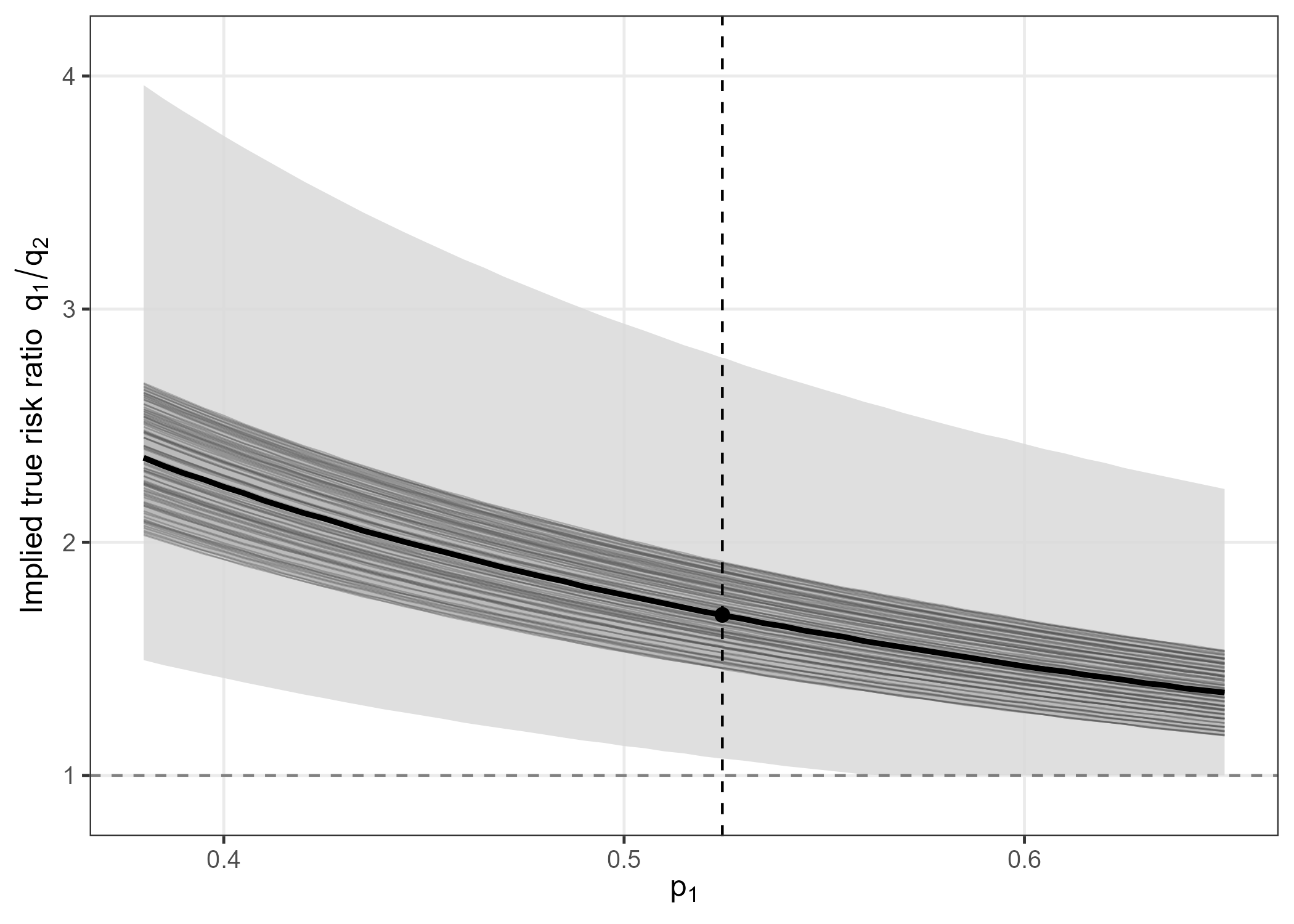}
\caption{Calibration-grid sensitivity analysis for the prostate illustration. Thin grey curves show 100 representative calibration settings from the grid \(q_2\in[0.02,0.04]\), \(p_2\in[0.33,0.43]\). The solid black curve shows the median implied true risk ratio over the full calibration grid. The darker shaded band shows the 2.5\%-97.5\% range of implied point estimates over the grid. The lighter shaded band shows the calibration-uncertainty interval, obtained from admissible grid values for which \(1.7\le RR^*\le3.3\). The vertical dashed line marks \(p_1\approx0.52\), corresponding to the empirical PSA-testing gradient.}
\label{fig:prostate_inverse_grid}
\end{figure}

For each grid point \(q_2\in[0.02,0.04]\) and \(p_2\in[0.33,0.43]\), we repeat the inverse calculation across values of \(p_1\). Figure~\ref{fig:prostate_inverse_grid} shows the resulting calibration-grid analysis. Thin grey curves show representative calibration settings. The solid black curve shows the median implied true risk ratio over the grid. The darker shaded band shows the 2.5\%--97.5\% range of implied point estimates over the grid. The lighter shaded band summarizes the corresponding conditional intervals across the calibration grid.

The calibration-grid analysis gives a broader range than the fixed-calibration calculation, but the qualitative conclusion is unchanged. At the empirical screening-gradient calibration \(p_1\approx0.52\), the median implied true risk ratio remains approximately \(1.69\), with a calibration-uncertainty interval of approximately \([1.07,2.79]\). Thus, allowing \(q_2\) and \(p_2\) to vary over the chosen calibration region does not remove the main conclusion: under plausible differential detection, the observed family-history association is compatible with a substantially smaller true risk ratio. This interval reflects uncertainty in the external calibration inputs and in the reported association endpoints, and should not be interpreted as a formal confidence interval for the biological family-history effect.
\section{Discussion}
\label{sec:discussion}

We used Markov models to study a risk-bias feedback loop generated by observed-history covariates. When true risk does not depend on history, uniform detection preserves the null risk ratio, while differential detection can create an apparent history effect. When true risk does depend on history, incomplete detection can attenuate the observed risk ratio by moving individuals with recent true history into less recent observed-history states. Targeted detection of putative risk groups can inflate the observed risk ratio. The simulations show the same point more broadly: detection regimes can distort both the observed risk ratio and the observed distribution of history. A regime that improves estimation of one parameter can worsen the estimation of the other.

The inverse sensitivity calculation makes the model a practical tool. Starting from an observed association, a baseline event probability, and plausible detection probabilities, it yields the true history effect implied by each detection scenario. Reported intervals for the observed association can be incorporated into the inverse calculation to reflect uncertainty in the sensitivity analysis.
Uncertainty in the external calibration inputs can be examined using a deterministic calibration grid. In our illustration, stronger detection among men with a known family history reduces the true risk ratio required to reproduce the observed association. This does not establish the presence of detection bias, but it makes the necessary detection assumptions explicit and shows how sensitive the implied true effect is to the model parameters.

The model is simple. It assumes fixed event and detection probabilities, no false positives, and a stable monitoring regime. The inverse calculation also relies on the stationary distribution of the Markov chain. This is a useful approximation when the history process and the monitoring regime have been operating for a long period, or when the target is a long-run population interpretation. It is less direct in settings with few generations, short follow-up, rapidly changing screening behavior, or new diagnostic technologies. In such cases, the stationary calculation should be interpreted as a long-run sensitivity benchmark rather than as a literal description of the observed study population. In practice, detection may change over calendar time, and the probability of observing an event may summarize many mechanisms: access to care, screening behavior, coding, registry capture, and communication. Some settings also involve intervention as well as observation, so the event process itself may change after detection. The inverse calculation also depends on external calibration values, which may come from different populations or time periods.

Future work could allow time-varying detection, use partial testing data to restrict plausible detection probabilities, or develop formal uncertainty models for combining published associations with external calibration inputs. The main message is that documented history should not be treated as an ordinary background covariate. When history is created by past detection and then used to guide future detection, the observed history effect combines true risk with monitoring.

\section*{Conflict of Interest}
The authors declare no conflicts of interest.

\section*{Data Availability Statement}
No individual-level data were analyzed in this study. The R code used to
reproduce the numerical analyses and figures is provided with the submission.

\newpage

\bibliographystyle{apalike}
\bibliography{references_observed_history}

\appendix

\section*{Appendix}

\section{Stationary distributions}
\label{app:stationary}

This appendix gives the algebra behind the Markov models in the main text. We use the notation of the manuscript. The true history state is denoted by \(H_t\), and the observed history state is denoted by \(H_t^*\). Both take values in \(\{1,\ldots,J+1\}\). 

\subsection{Equal true risk}
\label{app:equal}

\(H_t^*\) has transition matrix
\[
M =
\begin{array}{c|ccccccc}
 & 1 & 2 & 3 & 4 & \cdots & J & J+1 \\
\hline
1 & q p_1 & 1-q p_1 & 0 & 0 & \cdots & 0 & 0 \\
2 & q p_2 & 0 & 1-q p_2 & 0 & \cdots & 0 & 0 \\
3 & q p_3 & 0 & 0 & 1-q p_3 & \cdots & 0 & 0 \\
\vdots & \vdots & \vdots & \vdots & \vdots & \ddots & \vdots & \vdots \\
J & q p_J & 0 & 0 & 0 & \cdots & 0 & 1-q p_J \\
J+1 & q p_{J+1} & 0 & 0 & 0 & \cdots & 0 & 1-q p_{J+1}
\end{array}.
\]
Rows not shown in full follow the same pattern.

Let \(\pi^*_j\) denote the stationary probability of observed state \(j\). For \(j=2,\ldots,J\), the only incoming transition to state \(j\) is from state \(j-1\) without an observed event. Hence
\[
\pi^*_j=\pi^*_{j-1}(1-q p_{j-1}),
\qquad j=2,\ldots,J,
\]
and therefore
\[
\pi^*_j
=
\pi^*_1\prod_{k=1}^{j-1}(1-q p_k),
\qquad j=2,\ldots,J.
\]

For the terminal state \(J+1\),
\[
\pi^*_{J+1}
=
\pi^*_J(1-q p_J)+\pi^*_{J+1}(1-q p_{J+1}).
\]
Thus
\[
\pi^*_{J+1}
=
\frac{\pi^*_J(1-q p_J)}{q p_{J+1}}
=
\pi^*_1
\frac{\prod_{k=1}^{J}(1-q p_k)}
{q p_{J+1}}.
\]

Finally, the normalizing condition \(\sum_{j=1}^{J+1}\pi^*_j=1\) gives
\[
\pi^*_1
=
\left[
1+
\sum_{j=2}^{J}
\prod_{k=1}^{j-1}(1-q p_k)
+
\frac{\prod_{k=1}^{J}(1-q p_k)}
{q p_{J+1}}
\right]^{-1}.
\]
The stationary probabilities $\pi^*_1,\ldots,\pi^*_{J+1}$ follow from the expressions above.

\subsection{History-dependent true risk}
\label{app:unequal}

Let
\[
q_i=\Pr(A_{t+1}=1\mid H_t=i),
\qquad
p_j=\Pr(A_{t+1}^*=1\mid A_{t+1}=1,H_t^*=j).
\]
Under perfect specificity, the observed history cannot be more recent than the true history, so the state space of the joint chain is
\[
\mathcal{S}=\{(i,j):1\leq i\leq j\leq J+1\}.
\]
For \((i,j),(k,\ell)\in\mathcal{S}\), the transition probabilities are
\[
M_{(i,j),(k,\ell)}
=
\begin{cases}
q_i p_j,
& (k,\ell)=(1,1),\\[5pt]
q_i(1-p_j),
& (k,\ell)=\bigl(1,\min\{j+1,J+1\}\bigr),\\[5pt]
1-q_i,
& (k,\ell)=\bigl(\min\{i+1,J+1\},\min\{j+1,J+1\}\bigr),\\[5pt]
0,
& \text{otherwise}.
\end{cases}
\]
Let \(K=|\mathcal{S}|=(J+1)(J+2)/2\), and order the states of \(\mathcal{S}\) lexicographically. Let \(\pi\) be the \(K\)-vector of stationary probabilities in this order. The stationarity equations are
\[
M^T\pi=\pi,
\qquad
\mathbf{1}^T\pi=1.
\]
Equivalently,
\[
(M^T-I_K)\pi=0,
\qquad
\mathbf{1}^T\pi=1.
\]
To compute \(\pi\), define \(B=M^T-I_K\). Let \(B^{\dagger}\) be the matrix obtained from \(B\) by replacing its last row by \(\mathbf{1}^T\), and let
\[
b=(0,\ldots,0,1)^T.
\]
Then the stationary distribution is obtained from
\[
B^{\dagger}\pi=b,
\qquad\text{that is,}\qquad
\pi=(B^{\dagger})^{-1}b.
\]
This gives the full stationary distribution of the joint chain.

One component has a simpler interpretation. Let \(\alpha_i=\Pr(H_t=i)\) be the stationary distribution of the true history chain. The true history process is itself a Markov chain, with transition probabilities determined only by \(q_1,\ldots,q_{J+1}\), not by the detection probabilities. For every true-history state \(i\),
\[
\alpha_i=\sum_{j=i}^{J+1}\pi_{ij}.
\]
When \(i=J+1\), the only possible joint state is \((J+1,J+1)\). Therefore
\[
\pi_{J+1,J+1}=\alpha_{J+1}.
\]
Thus, the stationary mass at \((J+1,J+1)\) depends only on the true event probabilities \(q_1,\ldots,q_{J+1}\), and not on the detection probabilities \(p_1,\ldots,p_{J+1}\).

\subsection{The two-state case \texorpdfstring{\((J=1)\)}{(J=1)}}
\label{app:J1}

When \(J=1\), the history state has two levels. State \(1\) denotes a recent event and state \(2\) denotes no recent event. The joint state space is
\[
\{(1,1),(1,2),(2,2)\}.
\]
The transition matrix, with rows and columns ordered as \((1,1),(1,2),(2,2)\), is
\[
M=
\begin{pmatrix}
q_1p_1 & q_1(1-p_1) & 1-q_1\\
q_1p_2 & q_1(1-p_2) & 1-q_1\\
q_2p_2 & q_2(1-p_2) & 1-q_2
\end{pmatrix}.
\]

Let $\pi=(\pi_{11},\pi_{12},\pi_{22})$ denote the stationary distribution. From the stationary equation for \(\pi_{22}\),
\[
\pi_{22}
=
(1-q_1)\pi_{11}
+
(1-q_1)\pi_{12}
+
(1-q_2)\pi_{22}.
\]
Therefore
\begin{equation}
q_2\pi_{22}
=
(1-q_1)(\pi_{11}+\pi_{12}).
\label{eq:app_j1_pi22_balance}
\end{equation}

From the stationary equation for \(\pi_{11}\),
\[
\pi_{11}
=
q_1p_1\pi_{11}
+
q_1p_2\pi_{12}
+
q_2p_2\pi_{22}.
\]
Using \eqref{eq:app_j1_pi22_balance},
\[
\pi_{11}
=
q_1p_1\pi_{11}
+
q_1p_2\pi_{12}
+
(1-q_1)p_2(\pi_{11}+\pi_{12}).
\]
Collecting terms gives
\[
\pi_{11}\{1-q_1p_1-(1-q_1)p_2\}
=
p_2\pi_{12}.
\]
Let
\[
D:=1-q_1p_1-(1-q_1)p_2 =
q_1(1-p_1)+(1-q_1)(1-p_2).
\]
Under \(0<q_1<1\) and \(0\leq p_2\leq p_1\leq1\), we have \(D\geq0\), with equality only in the complete-detection case \(p_1=p_2=1\). For \(D>0\),

\begin{equation}
\pi_{11}
=
\frac{p_2}{D}\pi_{12}.
\label{eq:app_j1_pi11_pi12}
\end{equation}

Using \eqref{eq:app_j1_pi22_balance} and \eqref{eq:app_j1_pi11_pi12},
\[
\pi_{22}
=
\frac{1-q_1}{q_2}(\pi_{11}+\pi_{12})
=
\frac{1-q_1}{q_2}
\left(
\frac{p_2}{D}\pi_{12}
+
\pi_{12}
\right).
\]
Substituting this expression and \eqref{eq:app_j1_pi11_pi12} into
\[
\pi_{11}+\pi_{12}+\pi_{22}=1
\]
gives
\[
\frac{p_2}{D}\pi_{12}
+
\pi_{12}
+
\frac{1-q_1}{q_2}
\left(
\frac{p_2}{D}\pi_{12}
+
\pi_{12}
\right)
=
1.
\]
Hence
\[
\left(1+\frac{p_2}{D}\right)
\left(1+\frac{1-q_1}{q_2}\right)
\pi_{12}
=
1,
\]
so
\[
\pi_{12}
=
\frac{1}
{\left(1+\frac{p_2}{D}\right)
\left(1+\frac{1-q_1}{q_2}\right)}.
\]
Since
\[
1+\frac{p_2}{D}
=
\frac{D+p_2}{D}
=
\frac{1+q_1(p_2-p_1)}{D}
\]
and
\[
1+\frac{1-q_1}{q_2}
=
\frac{1-q_1+q_2}{q_2},
\]
we obtain
\[
\pi_{12}
=
\frac{q_2D}
{(1-q_1+q_2)\{1+q_1(p_2-p_1)\}}.
\]
Substituting the definition of \(D\) gives
\[
\pi_{12}
=
\frac{q_2\{1-q_1p_1-p_2(1-q_1)\}}
{(1-q_1+q_2)\{1+q_1(p_2-p_1)\}}.
\]
Then
\[
\pi_{11}
=
\frac{p_2}{D}\pi_{12}
=
\frac{q_2p_2}
{(1-q_1+q_2)\{1+q_1(p_2-p_1)\}},
\]
and
\[
\pi_{22}
=
\frac{1-q_1}{q_2}(\pi_{11}+\pi_{12})
=
\frac{1-q_1}{1-q_1+q_2}.
\]
Therefore
\begin{equation}
\boxed{
\begin{aligned}
\pi_{22}
&=
\frac{1-q_1}{1-q_1+q_2},\\[6pt]
\pi_{12}
&=
\frac{q_2\{1-q_1p_1-p_2(1-q_1)\}}
{(1-q_1+q_2)\{1+q_1(p_2-p_1)\}},\\[6pt]
\pi_{11}
&=
\frac{q_2p_2}
{(1-q_1+q_2)\{1+q_1(p_2-p_1)\}}.
\end{aligned}
}
\label{eq:app_j1_stationary}
\end{equation}

The first expression shows that \(\pi_{22}\) does not depend on the detection probabilities. This is because \((2,2)\) is the terminal state for the true process when \(J=1\), and its stationary mass is determined by the true transition probabilities.

\section{Proof of Proposition~\ref{prop:j1_rr}}
\label{app:RRJ1}

For the two-state model, substituting the stationary probabilities from
Appendix~\ref{app:J1} into Equation~\eqref{eq:j1_rr} gives
\begin{equation}
RR^*
=
\frac{q_1p_1}
{q_2p_2(1-q_1p_1)}
\left[
(1-q_1+q_2)\{1+q_1(p_2-p_1)\}
-
q_2p_2
\right].
\label{eq:app_j1_rr_closed}
\end{equation}

\paragraph{Part 1.}
Suppose \(p_1=p_2=p<1\). From Equation~\eqref{eq:app_j1_rr_closed},
\(RR^*\leq RR\) is equivalent to
\[
\frac{q_1p}
{q_2p(1-q_1p)}
\left[
(1-q_1+q_2)-q_2p
\right]
\leq
\frac{q_1}{q_2}.
\]
After cancellation, this is equivalent to
\[
(1-q_1+q_2)-q_2p
\leq
1-q_1p,
\]
or
\[
(q_1-q_2)(1-p)\geq 0.
\]
Therefore,
\[
RR^*\leq RR.
\]
Equality holds only if \(p=1\) or \(q_1=q_2\). This proves Part 1.

\paragraph{Part 2.}
Suppose \(p_1=1\) and \(p_2<1\). From Equation~\eqref{eq:app_j1_rr_closed},
\(RR^*>RR\) is equivalent to
\[
(1-q_1+q_2)(1-q_1+q_1p_2)-q_2p_2
>
p_2(1-q_1).
\]
The difference between the left-hand side and the right-hand side is
\[
\begin{aligned}
\Delta
&=
(1-q_1+q_2)(1-q_1+q_1p_2)
-
q_2p_2
-
p_2(1-q_1)\\
&=
(1-q_1)(1-q_1+q_2)(1-p_2).
\end{aligned}
\]
Since \(q_1<1\) and \(p_2<1\), we have \(\Delta>0\), and hence
\[
RR^*>RR.
\]
This proves Part 2.

\paragraph{Part 3.}
We show that \(RR^*\) is increasing in \(p_1\), for fixed \(q_1,q_2\), and \(p_2\). Write Equation~\eqref{eq:app_j1_rr_closed} as
\begin{equation}
RR^*
=
C
\frac{p_1\{B+A(p_2-p_1)\}}
{1-q_1p_1},
\label{eq:app_j1_rr_ABC}
\end{equation}
where
\[
A=q_1(1-q_1+q_2),
\qquad
B=1-q_1+q_2-q_2p_2,
\qquad
C=\frac{q_1}{q_2p_2}.
\]
Differentiating Equation~\eqref{eq:app_j1_rr_ABC} with respect to \(p_1\) gives
\[
\frac{\partial RR^*}{\partial p_1}
=
\frac{C}
{(1-q_1p_1)^2}
\left[
B+Ap_2-2Ap_1+Aq_1p_1^2
\right].
\]
The expression in brackets is decreasing in \(p_1\), because its derivative is
\[
-2A(1-q_1p_1)<0.
\]
Therefore, on the interval \(p_1\leq 1\), it is bounded below by its value at \(p_1=1\).  Substituting \(p_1=1\) into the expression in brackets gives:
\[
\begin{aligned}
B+Ap_2-2A+Aq_1
&=
1-q_1+q_2-q_2p_2
+
q_1p_2(1-q_1+q_2)\\
&\quad
-
2q_1(1-q_1+q_2)
+
q_1^2(1-q_1+q_2)\\
&=
(1-q_1)
\left[
(1-q_1+q_2)(1-q_1)
+
p_2(q_1-q_2)
\right].
\end{aligned}
\]
Under \(q_1\geq q_2\), this expression is positive. Therefore
\[
\frac{\partial RR^*}{\partial p_1}>0.
\]
This proves Part 3.

\paragraph{Part 4.}
Assume \(p_2<1\) and \(q_1>q_2\). By Part 1, at \(p_1=p_2\),
\[
RR^*<RR.
\]
By Part 2, at \(p_1=1\),
\[
RR^*>RR.
\]
By Part 3, \(RR^*\) is continuous and strictly increasing in \(p_1\). Therefore, there is a unique value
\[
p_1^*\in(p_2,1)
\]
such that
\[
RR^*=RR.
\]

To obtain the explicit threshold, set \(RR^*=RR\) in Equation~\eqref{eq:app_j1_rr_ABC}. This gives
\[
C
\frac{p_1\{B+A(p_2-p_1)\}}
{1-q_1p_1}
=
\frac{q_1}{q_2}.
\]
Using \(C=q_1/(q_2p_2)\), this is equivalent to
\[
p_1\{B+A(p_2-p_1)\}
=
p_2(1-q_1p_1),
\]
or
\[
Ap_1^2-(B+Ap_2+q_1p_2)p_1+p_2=0.
\]
The two algebraic roots are
\[
p_1
=
\frac{
B+Ap_2+q_1p_2
\pm
\sqrt{(B+Ap_2+q_1p_2)^2-4Ap_2}
}
{2A}.
\]
By uniqueness, exactly one of these roots lies in \((p_2,1)\). That root is \(p_1^*\). This proves Part 4.

\paragraph{Part 5.}
Finally, recall that
\[
\pi_{11}
=
\frac{q_2p_2}
{(1-q_1+q_2)\{1+q_1(p_2-p_1)\}}.
\]
Differentiating with respect to \(p_2\) gives
\[
\frac{\partial \pi_{11}}{\partial p_2}
=
\frac{q_2(1-q_1p_1)}
{(1-q_1+q_2)\{1+q_1(p_2-p_1)\}^2}
>0,
\]
since \(q_1p_1<1\). 

Moreover, the parameter \(p_1\) appears only in the factor
\[
1+q_1(p_2-p_1)
\]
in the denominator. This factor is positive under \eqref{eq:qp_conditions} and decreases as \(p_1\) increases. Hence, \(\pi_{11}\) is increasing in \(p_1\). This proves Part 5.

\end{document}